\documentclass[12pt,a4paper]{elsarticle}
\makeatletter
\def\ps@pprintTitle{%
	\let\@oddhead\@empty
	\let\@evenhead\@empty
	\def\@oddfoot{\centerline{\thepage}}%
	\let\@evenfoot\@oddfoot}
\makeatother

\usepackage{amssymb}
\usepackage{amsthm}
\usepackage{amsmath} 
\usepackage{gensymb}
\usepackage{bm}

\usepackage{float}
\restylefloat{table}

\usepackage{graphicx}
\usepackage{caption}
\usepackage{subcaption}
\graphicspath{{Figures/}}

\usepackage{tikz}
\usetikzlibrary{shapes.geometric, arrows}
\tikzstyle{highlevel} = [rectangle, minimum width=3cm, minimum height=1cm, text centered, text width=3cm, draw=black, text=black, fill=cyan!30]
\tikzstyle{lowlevel} = [rectangle, minimum width=2.5cm, minimum height=1cm, text centered, text width=2.5cm, draw=black, text=black, fill=red!30]
\tikzstyle{inputoutput1} = [trapezium, trapezium left angle=70, trapezium right angle=110, minimum width=2.5cm, minimum height=1cm, text width=1.25cm, text centered, draw=black, text=black, fill=green!30]
\tikzstyle{inputoutput2} = [trapezium, trapezium left angle=70, trapezium right angle=110, minimum width=2.5cm, minimum height=1cm, text centered, text width=2.5cm, draw=black, text=black, fill=green!30]
\tikzstyle{external} = [rectangle, rounded corners, minimum width=2.5cm, minimum height=1cm, text centered, text width=2.5cm, draw=black, text=black, fill=magenta!30]
\tikzstyle{arrow} = [thick,->,>=stealth]

\usepackage[T1]{fontenc} 
\usepackage{txfonts} 
\usepackage{xcolor} 
\usepackage{hyperref}  

\biboptions{sort&compress}

\begin{document}
	
\begin{frontmatter}
	
	
	
	\title{\texttt{Exasim}\tnoteref{label1}: Generating Discontinuous Galerkin Codes for Numerical Solutions of Partial Differential Equations on Graphics Processors}
	 \tnotetext[label1]{\url{https://github.com/exapde/Exasim} }
	
	
	\author{Jordi Vila-P\'erez\corref{cor1}}
    \cortext[cor1]{Corresponding author.}
	\ead{jvilap@mit.edu}
	\author{R. Loek Van Heyningen}
	\author{Ngoc-Cuong Nguyen}
	\author{Jaume Peraire}
	\address{Department of Aeronautics and Astronautics, Massachusetts Institute of Technology, Cambridge, 02139, Massachusetts, USA}

	\begin{abstract}	
		This paper presents an overview of  the functionalities and applications of \texttt{Exasim}, an open-source code for generating high-order discontinuous Galerkin codes to numerically solve parametrized partial differential equations (PDEs).
		The software combines high-level and low-level languages to construct parametrized PDE models via Julia, Python or Matlab scripts and produce high-performance C++ codes for solving the PDE models on CPU and Nvidia GPU processors with distributed memory.
		\texttt{Exasim} provides matrix-free discontinuous Galerkin discretization schemes together with scalable reduced basis preconditioners and Newton-GMRES solvers, making it suitable for accurate and efficient approximation of wide-ranging classes of PDEs.
	\end{abstract}
	
	
\end{frontmatter}


\section{Motivation and significance}
\label{sc:motivation}

The use of high-order methods to solve partial differential equations (PDEs) has experienced a growing interest among practitioners \cite{Kroll2015,Kroll2009,Wang2013} in different areas of engineering and science.
Their increased accuracy at a reduced computational cost and their low diffusion and dispersion errors confer them a major advantage when compared to low-order schemes \cite{Slotnick2014,Ekaterinaris2005}.
In particular, DG formulations have become one of the most adopted high-order approaches in many different areas \cite{Cockburn-CKS:2000,Hesthaven2010,Cockburn-CLS:1989,Cockburn-CS:1998,Bassi-BR:97,Bassi-BR:1997b,Bassi-BR:2002}.
DG methods rely on a locally conservative formulation that ensures high-order accuracy on unstructured meshes.
In addition, they provide a stable definition of the convection operator and allow suitable $hp$-adaptivity strategies \cite{HartmannHouston:2003,Balan-BWM:2015,Giorgiani-GFH:2014,Cangiani2017} and an efficient exploitation of parallel computing architectures \cite{Roca-RNP:2013}.

Because of these reasons, different classes of DG discretizations have been proposed over the last years \cite{Cockburn-CG:2004,Cockburn2009,Nguyen2015c,CockburnShu1998,Cockburn-CNP:2010,PerairePersson08}.
In particular, hybridized DG (HDG) methods have gained popularity for the numerical solution of all classes of PDEs \cite{VilaPerez2020,RS-SH:16,Ciuca2020,Fernandez2018a,Fernandez2017a,Nguyen2012,Nguyen-NPC:20011acoustics,Giacomini-GKSH:2018,Sevilla-SGKH:2018,VidalCodina2018,VidalCodina2021}, based on their superiority with respect to other DG alternatives both in terms of accuracy and computational complexity \cite{Huerta-HARP:2013,May-WBMS-14}.
However, their conjunction with nonlinear solvers imposes a high memory footprint, either in terms of building the system matrices, or dealing with larger systems of equations in matrix-free approaches \cite{Kronbichler2019}.
As a result, these methods offer a poorer scalability on GPU platforms, what makes them unsuitable for solving large problems \cite{Terrana2020}.

This paper presents \texttt{Exasim}, an open-source software for generating high-order DG codes, based on the local DG (LDG) method \cite{CockburnShu1998,CockburnKanschatSchoetzauNS05}.
\texttt{Exasim} performs an implicit matrix-free approach \cite{Terrana2020,Nguyen2022} that makes the code suitable for running on multiple architectures using both CPUs and GPUs, allowing the solution of large-scale problems.
Moreover, it exploits a parametrized formulation of PDEs, expressing them as first order systems of equations with the eventual inclusion of ordinary differential equations (ODEs) to form a system of differential-algebraic equations (DAEs).
This general framework reduces the mathematical description of the model to the definition of state variables, fluxes, source terms, together with initial and boundary conditions, and allows the user to model multiple PDE systems.
Furthermore, the software presents a user interface in a high-level language (Julia, Python or Matlab) where one can specify the aforementioned terms symbolically in a seamless effort.
Then, the code performs a preprocessing stage that generates C++ code, which interfaces with C++ and CUDA kernels and allows the software to run on different computing platforms.

With this approach, \texttt{Exasim} offers an open-source product that can be easily adopted for users with any kind of expertise in DG discretizations.
At the same time, it serves as an advanced research tool, capable to run in different architectures with multiple processors and with suitable scalability properties, thus allowing to tackle complex problems that are beyond the capabilities of existing codes \cite{Nguyen2022,Terrana2020}.
Indeed, whereas the implicit-in-time matrix-free approach shares some common features with some available DG codes \cite{ExaDG2020,Arndt2021,Anderson2021}, it contrasts with most of the available open-source DG approaches, such as \cite{flexi_general,Cantwell2015,Ching2022,Giacomini2020,Dedner2010,NGSolve,Prudhomme2021,Reuter2020}, which are CPU-based, and also with other codes running on GPUs which rely on explicit time-marching schemes, like \cite{Kloeckner2012,Witherden2014}.

The remainder of this paper is structured as follows.
Section~\ref{sc:PDEmodel} describes the parametrized PDE models and discretization methods that \texttt{Exasim} handles.
Then, section~\ref{sc:software} details the code architecture and its different functionalities.
A number of examples are presented in section~\ref{sc:examples} to illustrate the capabilities of the software.
Finally, sections~\ref{sc:impact} and \ref{sc:conclusions} summarize the impact of this work and the main conclusions of this paper, respectively.

\section{Models and Discretization Methods} \label{sc:PDEmodel}
\texttt{Exasim} produces executable DG code to solve a wide variety of PDE models that can be described under general parametrized formulations and classified under convection, diffusion or wave-type equations (models C, D and W in the software), such as those listed in Table~\ref{tb:models}, and described as follows.
\begin{table}[H]
	\begin{tabular}{|p{3.7cm}|p{9.25cm}|}
		\hline
		\textbf{Convection model} & Linear and nonlinear convection, Burgers equation, Euler equations, shallow water equations.\\
		\hline
		\textbf{Diffusion model} & Poisson equation, convection-diffusion equations, linear and nonlinear elasticity, compressible and incompressible Navier-Stokes.\\
		\hline
		\textbf{Wave model} & Wave equation, linear and nonlinear elastodynamics, Maxwell's equations.\\
		\hline
	\end{tabular}
	\caption{Examples of PDEs belonging to each of the model categories in \texttt{Exasim}.}
	\label{tb:models} 
\end{table}

\subsection{Parametrized PDE models} 
The underlying PDE system must be written as a set of first-order PDEs and can be coupled as well with a certain ordinary differential equation (ODE) to form a system of differential-algebraic equations.
For instance, the diffusion model, defined in the open domain $\Omega \subset \mathbb{R}^{n_d}$ for $t_f>0$, and expressed in its more general version, reads as
\begin{subequations} \label{eq:PDEmodel}
	\begin{alignat}{2}
		\bm q + \nabla \bm u & =  \bm 0,  & \quad \mbox{in } \Omega \times (0, t_f], \label{eq:mixed} \\
		\bm m(\bm{\tilde{u}}, \bm x,t, \bm \mu) \frac{\partial \bm{u}}{\partial t} + \nabla \cdot \bm{f}(\bm{\tilde{u}}, \bm x,t, \bm \mu) &=  \bm s(\bm{\tilde{u}},  \bm x,t, \bm \mu), & \quad \mbox{in } \Omega \times (0, t_f], \label{eq:primal} \\
		\alpha \frac{\partial \bm w}{\partial t}   +  \beta \bm w & =  \bm s_w(\bm \tilde{u}, \bm x, t, \bm \mu),  & \quad \mbox{in } \Omega \times (0, t_f], \label{eq:ode}
	\end{alignat}
with appropriate initial and boundary conditions. Here, the set of state variables $\bm{\tilde{u}} = (\bm u, \bm q, \bm w) \in \mathbb{R}^{n_{cu}} \times \mathbb{R}^{n_{cu} \times n_d} \times \mathbb{R}^{n_w}$ is the exact solution of the PDE model, $\bm x \in \Omega$ is the vector of coordinate variables, $t$ represents time variable in $(0, t_f]$, and $\bm \mu \in \mathbb{R}^{n_{param}}$ is a vector of physical parameters.
Additionally, the vector-valued function  $\bm m \in \mathbb{R}^{n_{cu}}$ is a mass function, the matrix-valued function $\bm f \in \mathbb{R}^{n_{cu}\times n_d}$ is a flux and the vector-valued function  $\bm s \in \mathbb{R}^{n_{cu}}$ is a source term. Similarly, $\alpha\in \mathbb{R}$, $\beta\in\mathbb{R}$ and $\bm s_w \in \mathbb{R}^{n_{w}}$ are, respectively, two parameters and a source term for the additional ODE.

The diffusion model (\ref{eq:mixed}--\ref{eq:primal}) reduces to the convection model when the state variables $\bm{\tilde{u}}$ do not contain $\bm q$ and equation~\ref{eq:mixed} is not included in the model. The wave model derives from the diffusion model when equation~\ref{eq:mixed} is replaced by
\begin{alignat}{2}
\frac{\partial \bm q}{\partial t} + \nabla \bm u  &=  \bm 0,   & \quad \mbox{in } \Omega \times (0, t_f],
\end{alignat}
and employs the ODE equation~\ref{eq:ode} to recover the displacement field, $\bm w$.
Finally, note that besides the diffusion, convection, and wave models, \texttt{Exasim} can solve higher-order PDE models by rewriting them as a first-order system of equations.    

\end{subequations}

\subsection{Discretization methods}
\texttt{Exasim} implements the LDG method for the spatial discretization and the diagonally implicit Runge-Kutta (DIRK) method for temporal discretization \cite{Alexander77}.
The LDG discretization of the PDE model \eqref{eq:PDEmodel} leads to a semi-discrete system of equations involving the source, flux, mass, numerical trace, and numerical flux functions.
In \texttt{Exasim}, these functions are input as mathematical expressions of symbolic variables in a script file.
This allows users to define the LDG discretization of a PDE model by merely writing mathematical functions in a high-level language setting.
Note as well that different DG methods can be implemented in \texttt{Exasim} for the spatial discretization of PDE systems by providing suitable expressions of the numerical traces and numerical fluxes.

On the other hand, \texttt{Exasim} employs several DIRK schemes for the time integration, from first-order implicit Euler to higher-order schemes, such as the three-stage four-order DIRK scheme.
The number of stages and order of accuracy can be specified by users.

\section{Software description} \label{sc:software}


\texttt{Exasim} combines a high-level interface for preprocessing and code generation with  C++ language to obtain high-performance codes that can run on both CPU and GPU architectures.
In particular, the functions and parameters that define the PDE model~\eqref{eq:PDEmodel} are specified via scripts in a high-level language (Julia, Python or Matlab).
An automatic code generation module is then responsible for converting them into C++ codes that handle fluxes, source terms, boundary conditions and initial conditions.
Finally, the kernel code, written in C++ with MPI-based parallelization and CUDA for GPUs, implements the corresponding discretization and solution methods. \texttt{Exasim} leverages a number of external libraries and software, such as the linear algebra libraries BLAS and LAPACK, Gmsh for mesh generation, METIS for mesh partitioning, GPU-aware MPI libraries, CUDA Toolkit for GPU architectures and Paraview for visualization. 

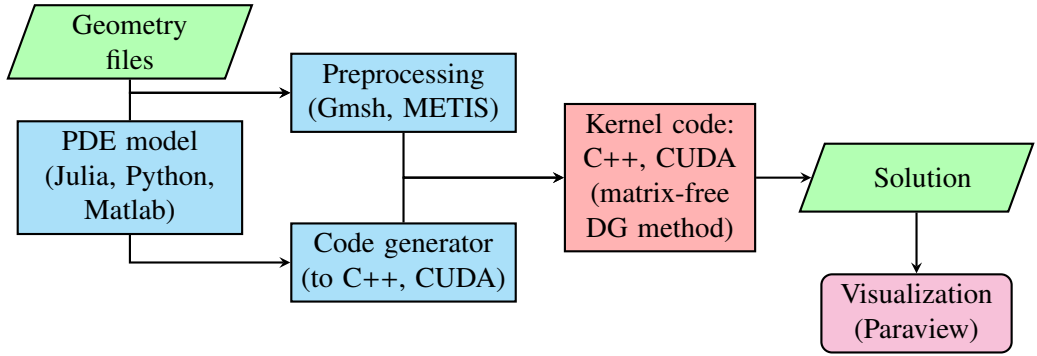
\begin{figure}[H]
	\begin{tikzpicture}[node distance=2cm, thick,scale=1, every node/.style={scale=0.9}]
		<TikZ code>
		\node (geom) [inputoutput2] {Geometry files};
		\node (pdemodel) [highlevel, below of=geom] {PDE model (Julia, Python, Matlab)};
		\node (prepro) [highlevel, right of=pdemodel, xshift=2cm,yshift=1.25cm] {Preprocessing (Gmsh, METIS)};
		\node (code) [highlevel, below of=prepro, yshift=-0.5cm] {Code generator (to C++, CUDA)};
		\node (kernel) [lowlevel, right of=pdemodel, xshift=5.75cm] {Kernel code: C++, CUDA (matrix-free DG method)};
		\node (solution) [inputoutput1, right of=kernel, xshift=1.75cm] {Solution};
		\node (visualization) [external, below of=solution] {Visualization (Paraview)};
		
		\draw [arrow] (geom) |- (prepro);
		\draw [arrow] (pdemodel) |- (prepro);
		\draw [arrow] (pdemodel) |- (code);
		\draw [arrow] (prepro) |- (kernel);
		\draw [arrow] (code) |- (kernel);
		\draw [arrow] (kernel) -- (solution);
		\draw [arrow] (solution) -- (visualization);
		
	\end{tikzpicture}
	\caption{Flowchart summarizing the code architecture of \texttt{Exasim} and the dependencies among the different processes. The color code is the following: processes implemented in high/low-level languages (blue/red, respectively), input and output files (green), external functionalities (magenta).}
	\label{fig:flowchart}
\end{figure}

A summary of this code architecture is depicted in Figure~\ref{fig:flowchart}, which illustrates the different tasks and dependencies together with some of the software main functionalities.
Some of the main capabilities of the code related to the solution methods, meshing and visualization, besides its GPU parallel performance are described as follows.

\subsection{Automatic code generation}

\texttt{Exasim} generates both standard C++ code and CUDA code from the mathematical functions written in Julia, Python, or Matlab. To this end, \texttt{Exasim} employs symbolic libraries such as Sympy and Matlab's symbolic toolbox to generate C code which is then converted to C++/CUDA code by \texttt{Exasim}'s code generator. The resulting code is automatically optimized, given that common subexpression elimination (CSE) tool is used to  eliminate duplicate expressions.

\subsection{High-order mesh generation}
\texttt{Exasim} provides a Mesh module to generate meshes for simple geometries.
Similarly, \texttt{Exasim} uses Gmsh \cite{Geuzaine2009} to generate meshes from geometry model files.
Nevertheless, any alternative open-source mesh generators such as CUBIT \cite{CUBIT}, CGAL \cite{CGAL}, DistMesh \cite{Persson2004},  TetGen \cite{Si2015}, Mmg \cite{Dapogny2014}, MeshLab \cite{Cignoni2008} or SALOME \cite{Nicolas2013}, among others, can be used for complex geometries.
Because a high-order mesh is needed for the DG discretization of a PDE model, \texttt{Exasim} produces the high-order mesh from a standard finite element mesh by curving its corresponding boundaries and mapping the nodes of the mesh appropriately \cite{Persson2009a,Gargallo-Peiro2015}.

\subsection{Jacobian-free Newton-Krylov solvers} \label{ssc:Discretization}
The LDG method is \texttt{Exasim}'s default discretization, since it enables an efficient implementation of a matrix-free solution method well suited for GPU architectures.
The resulting system of equations arising from DG/DIRK discretization is solved by using a matrix-free Newton-GMRES method, thus avoiding the need to construct Jacobian matrices. The performance of the GMRES solver is accelerated by means of a matrix-free preconditioner which is constructed using the reduced basis method and a low-rank approximation to the Jacobian matrix. The matrix-vector products in GMRES can be computed in two different ways in \texttt{Exasim}, either by means of a finite difference approximation or by an automatic differentiation (AD) approach relying on the external package Enzyme \cite{Moses2020,Moses2021}. In addition, a number of different algorithms, including tensor-product with sum-factorization for residual evaluation or an automatic tuning for customized GPU allocation, are used to optimize the code performance. More details of the different algorithms employed for the discretization and solution method can be found in \cite{Nguyen2022,Terrana2020}.

\subsection{Visualization}
\texttt{Exasim} uses Paraview to visualize and analyze the numerical solutions. To this end, a postprocessing tool is employed to generate the corresponding \emph{VTK/VTU} files from the solution data. The visualization is performed immediately once the simulation is completed.

\subsection{GPU scalability}
Whereas \texttt{Exasim} can run both on CPU and GPU architectures, the code features a set of numerical algorithms suited to optimize the performance in GPU systems.
Indeed, a significant performance gain for GPU architectures has been reported in \cite{Terrana2020}, indicating an improvement of more than one order of magnitude in runtime with respect to CPU machines.

The software presents excellent performance in weak and strong scaling tests, as illustrated in Table \ref{tb:scaling}, for the direct numerical simulation (DNS) of the Purdue flared cone \cite{Hader2019,Chynoweth2019}.
The simulation employed third-order DG and DIRK schemes and up to 768 nodes at the OLCF's Summit supercomputer, with one MPI rank per GPU.
%
A degradation of about 5\% is obtained in the weak scaling test when increasing from 24 to 768 nodes, whereas a slow degradation can be observed as well in the strong scaling results as the number of nodes increases.

\begin{table}[H]
	\begin{tabular}{|c|ccc|ccc|}
		\cline{2-7}
		\multicolumn{1}{c|}{}& \multicolumn{3}{c|}{\textbf{Weak scaling}} & \multicolumn{3}{c|}{\textbf{Strong scaling}}\\
		\hline
		\textbf{Nodes} & \textbf{DOFs} & \textbf{Time (s)} & \textbf{Time ratio} & \textbf{DOFs} & \textbf{Time (s)} & \textbf{Time ratio} \\
		\hline
		24 & 0.408B & 3.10 & 1.000 & 1.632B & 12.45 & 1.000\\
		48 & 0.816B & 3.12 & 1.006 & 1.632B & 6.25 & 0.502\\
		96 & 1.632B & 3.14 & 1.024 & 1.632B & 3.14 & 0.252\\
		192 & 3.264B & 3.17 & 1.023 & 1.632B & 1.59 & 0.128\\
		384 & 6.858B & 3.20 & 1.032 & 1.632B & 0.81 & 0.065\\
		768 & 13.056B & 3.25 & 1.048 & 1.632B & 0.43 & 0.035\\
		\hline
	\end{tabular}
	\caption{Weak scaling and strong scaling tests of \texttt{Exasim} on NVIDIA Tesla V100 GPUs. The time column indicates the physical time needed per time-step.} 
	\label{tb:scaling}
\end{table}

\section{Illustrative Examples} \label{sc:examples}

\texttt{Exasim} has a large collection of examples including convection-diffusion, heat transfer, compressible flows, wave propagation and magnetohydrodynamics problems. This section presents four different examples which are representatives of convection, diffusion and wave models.

\subsection{Convergence study on the Poisson equation}
A 3D analytical solution of the Poisson equation is presented to verify the convergence properties of the proposed numerical method, in particular for the diffusion model.
The example corresponds to a case of heat diffusion with a source term, i.e. $-\nabla^2 u = f$ on the unit cube $\Omega = (0,1)^3$, with analytical solution $u = \sin (\pi x) \sin (\pi y) \sin (\pi z)$.
The problem is solved in a set of structured tetrahedral grids and using different polynomial degrees of approximation, $p$.
%
The approximation errors in the different mesh refinements for both the primal and mixed variables, $u_h$ and $\bm{q_h}$, are detailed in Table~\ref{tb:convergence}. The expected optimal convergence rates of $p+1$ are obtained for the approximation of the primal variable, $u_h$, whereas convergence of order $p$ is obtained for the mixed variables, $\bm{q_h}$.
\begin{table}[H]
	\begin{tabular}{|c||cc|cc|cc|cc|}
		\hline
		\multicolumn{1}{|c||}{Mesh} & \multicolumn{2}{c|}{$p=1$} & \multicolumn{2}{c|}{$p=2$} & \multicolumn{2}{c|}{$p=3$} & \multicolumn{2}{c|}{$p=4$}\\
		\textbf{$1/n$} & Error & Rate & Error & Rate & Error & Rate & Error & Rate \\
		\hline
		\multicolumn{9}{|c|}{ $E_{\mathcal{L}_2}(u_h) =  \|u_h - u\|_{\mathcal{L}_2 (\Omega)} / \|u\|_{\mathcal{L}_2 (\Omega)}$ } \\
		\hline
		2 & 1.51e-01 & -- & 3.65e-02 & -- & 8.95e-03 & -- & 2.31e-03 & --\\
		4 & 5.68e-02 & 1.41 & 4.51e-03 & 3.02 & 6.99e-04 & 3.68 & 8.18e-05 & 4.82\\
		6 & 1.68e-02 & 1.76 & 5.69e-04 & 2.99 & 4.70e-05 & 3.90 & 2.65e-06 & 4.95\\
		16 & 4.44e-03 & 1.92 &7.28e-05 & 2.97 & 3.00e-06 & 3.97 & 8.45e-08 & 4.97\\
		32 & 1.13e-03 & 1.97 & 9.24e-06 & 2.98 & 1.91e-07 & 3.98 & 5.58e-09 & $\times$\\
		\hline
		\multicolumn{9}{|c|}{$E_{\mathcal{L}_2}(q_h) = \|\bm{q_h} - \bm{q}\|_{\mathcal{L}_2 (\Omega)}/\|\bm{q}\|_{\mathcal{L}_2 (\Omega)}$} \\
		\hline
		2 & 5.71e-01 & -- & 2.19e-01 & -- & 5.67e-02 & -- & 1.31e-02 & --\\
		4 & 3.67e-01 & 0.64 & 5.60e-02 & 1.97 & 9.47e-03 & 2.58 & 8.87e-04 & 3.89 \\
		6 & 2.05e-01 & 0.84 & 1.33e-02 & 2.07 & 1.33e-03 & 2.83 & 5.41e-05 & 4.03\\
		16 & 1.07e-01 & 0.94 &3.26e-03 & 2.03 & 1.74e-04 & 2.94 & 3.37e-06 & 4.01\\
		32 & 5.42e-02 & 0.98 & 8.12e-04 & 2.01 & 2.20e-05 & 2.98 & 7.48e-07 & $\times$\\
		\hline
	\end{tabular}
	\caption{Poisson example -- History of convergence of the primal and mixed variables, $u_h$ and $\bm{q_h}$ (top and bottom respectively) in the 3D Poisson example, using uniform meshes of tetrahedrons and different polynomial degrees of approximation. $\times$ indicates that convergence rates cannot be accurately computed due to finite precision arithmetic issues.} 
	\label{tb:convergence}
\end{table}

\subsection{Scattering of a planar wave by a circular cylinder}
\begin{figure}[H]
	\centering
	\begin{subfigure}[b]{0.28\textwidth}
		\centering
		\includegraphics[width=\textwidth]{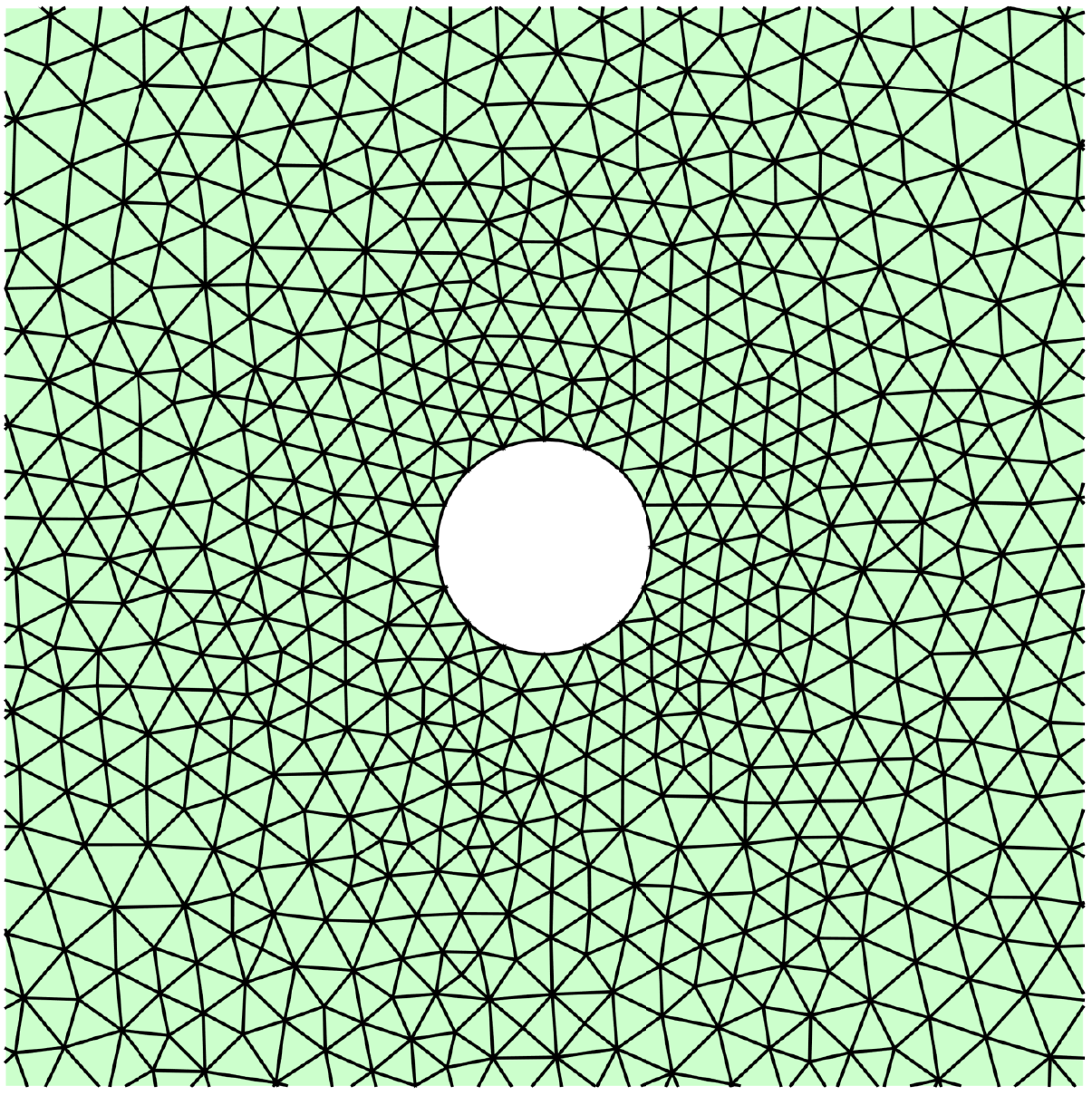}
		\caption{Close up view of the mesh.}
		\label{fig:mesh}
	\end{subfigure}
	\hfill
	\begin{subfigure}[b]{0.34\textwidth}
		\centering
		\includegraphics[width=\textwidth]{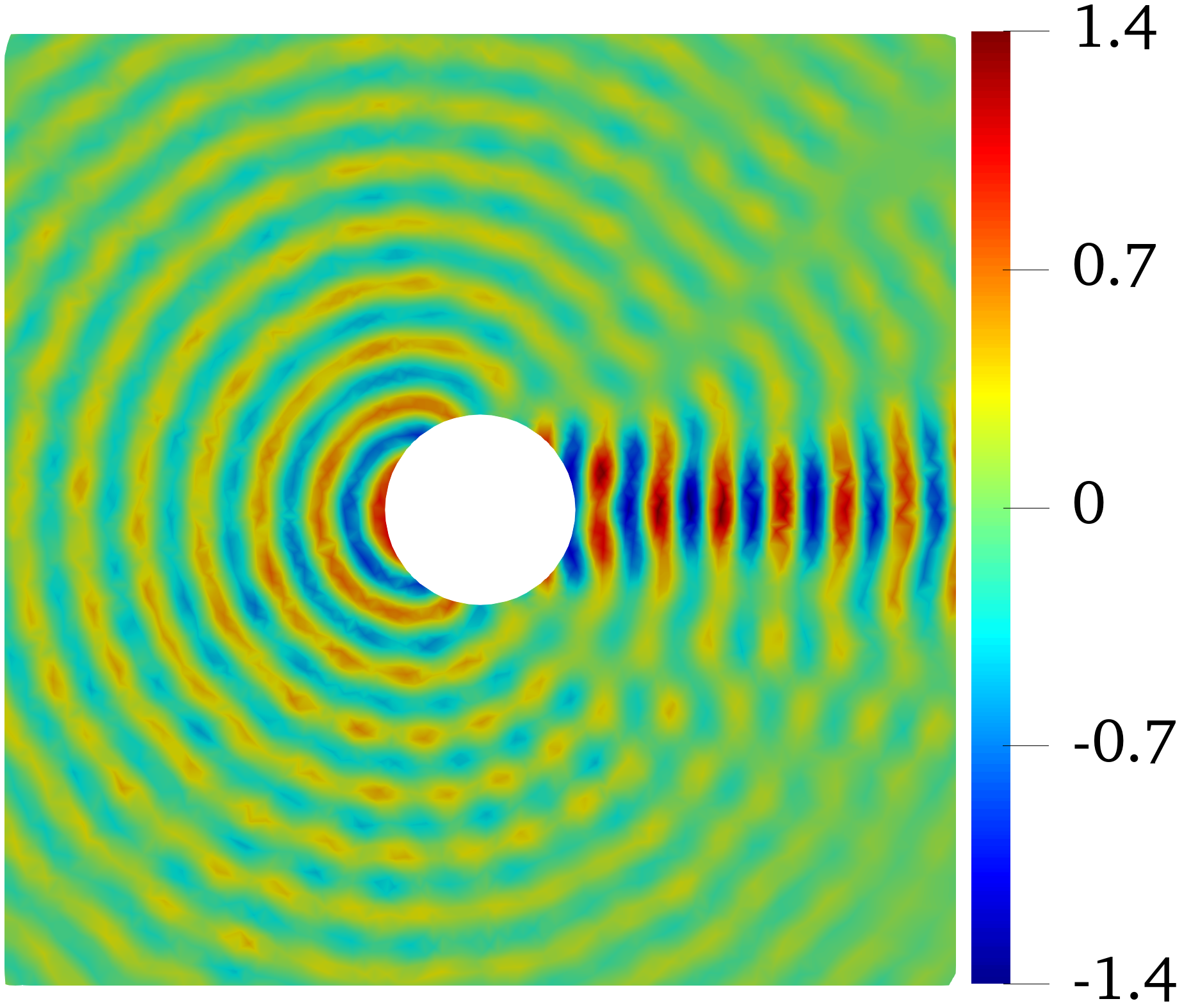}
		\caption{Displacement ($p=3$).}
		\label{fig:up3}
	\end{subfigure}
	\begin{subfigure}[b]{0.34\textwidth}
    	\centering
    	\includegraphics[width=\textwidth]{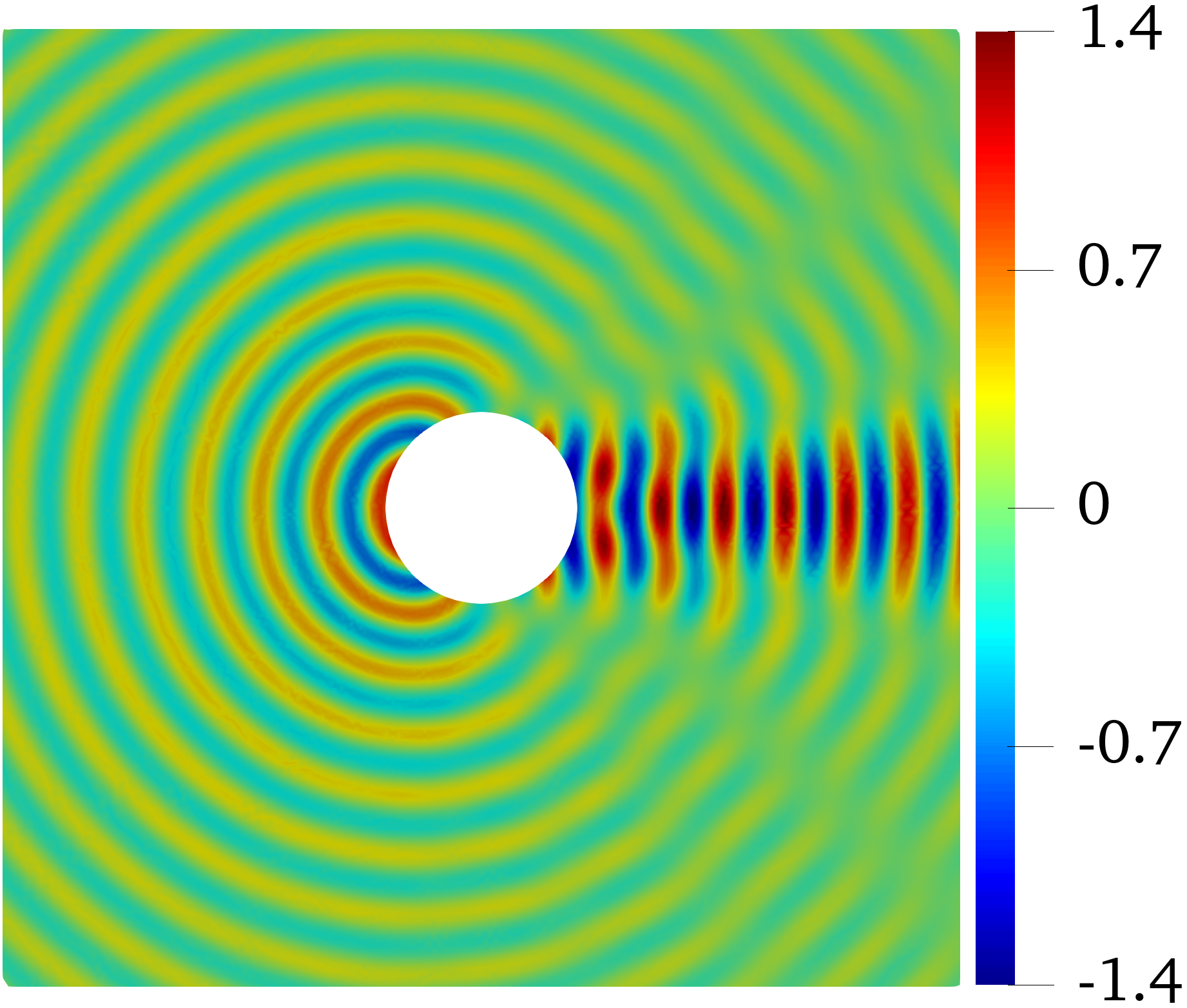}
    	\caption{Displacement ($p=5$).}
    	\label{fig:up5}
    \end{subfigure}
	\begin{subfigure}[b]{0.45\textwidth}
		\centering
		\includegraphics[width=\textwidth]{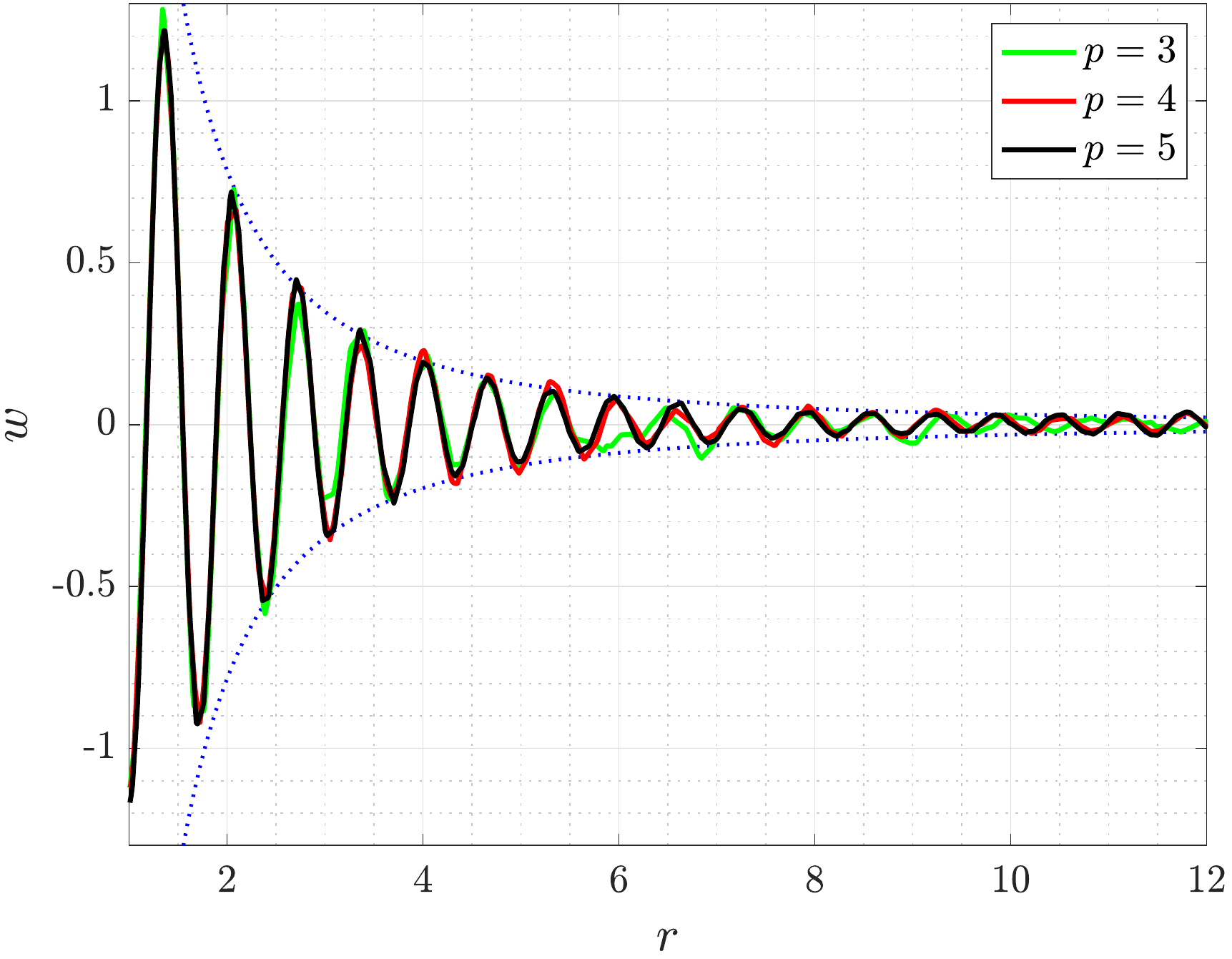}
		\caption{Displacement along $\theta = \pi/9$.}
		\label{fig:decay}
	\end{subfigure}
	\begin{subfigure}[b]{0.45\textwidth}
		\centering
		\includegraphics[width=\textwidth]{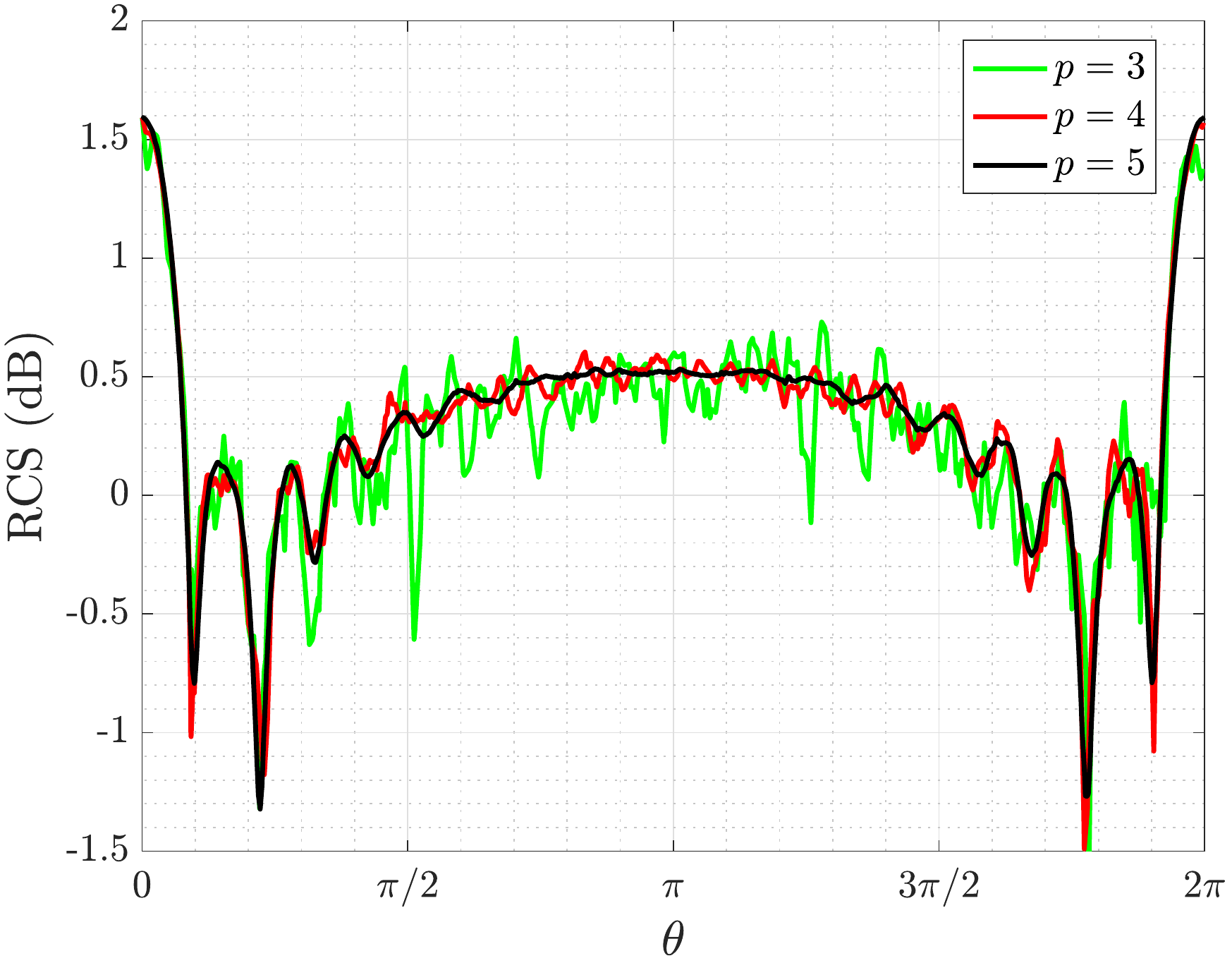}
		\caption{Radar cross-section.}
		\label{fig:RCS}
	\end{subfigure}
	\caption{Wave scattering -- Detail of (a) the unstructured mesh with curved boundaries, the instantaneous displacement field at $t=50T$ for (b) $p=3$, (c) $p=5$ and (d) along the line $\theta = \pi/9$, and (e) the radar cross-section. The dotted lines in (d) indicate a quadratic decay.}
	\label{fig:wavescattering}
\end{figure}

The second example presented in this work corresponds to the propagation of an acoustic planar wave of wavenumber $\bm{k} = (10,0)$ in a medium with unit permittivity, $\varepsilon = 1$, and speed of sound, $c=1$, scattered by a 2D cylinder of unitary radius.
The problem is solved in the square domain $\Omega = (-12,12)^2$, with absorbing boundary conditions on the outer boundaries and Neumann boundary conditions on the cylinder boundary. The numerical simulation is performed using different polynomial orders of approximation, from $p=3$ to $p=5$, and a DIRK(3,4) scheme, employing an unstructured mesh composed by 4224 triangles with curved boundaries, shown in Figure~\ref{fig:mesh}.
The scattered wave solution is depicted in Figures~\ref{fig:up3} and~\ref{fig:up5} for $p=3$ and $p=5$, respectively, after 50 periods of time, whereas Figure~\ref{fig:decay} illustrates the quadratic decay of the displacement field along the line $\theta = \pi/9$.
Finally, Figure~\ref{fig:RCS} shows the radar cross-section (RCS)~\cite{Baumeister1994} for different orders of approximation, measuring the propagated energy and highlighting the importance of high-order approximations in time and space to ensure low dispersion and diffusion errors on the wave propagation.

\subsection{Bickley jet}
\begin{figure}[H]
	\centering
	\begin{subfigure}[b]{0.29\textwidth}
		\centering
		\includegraphics[width=\textwidth]{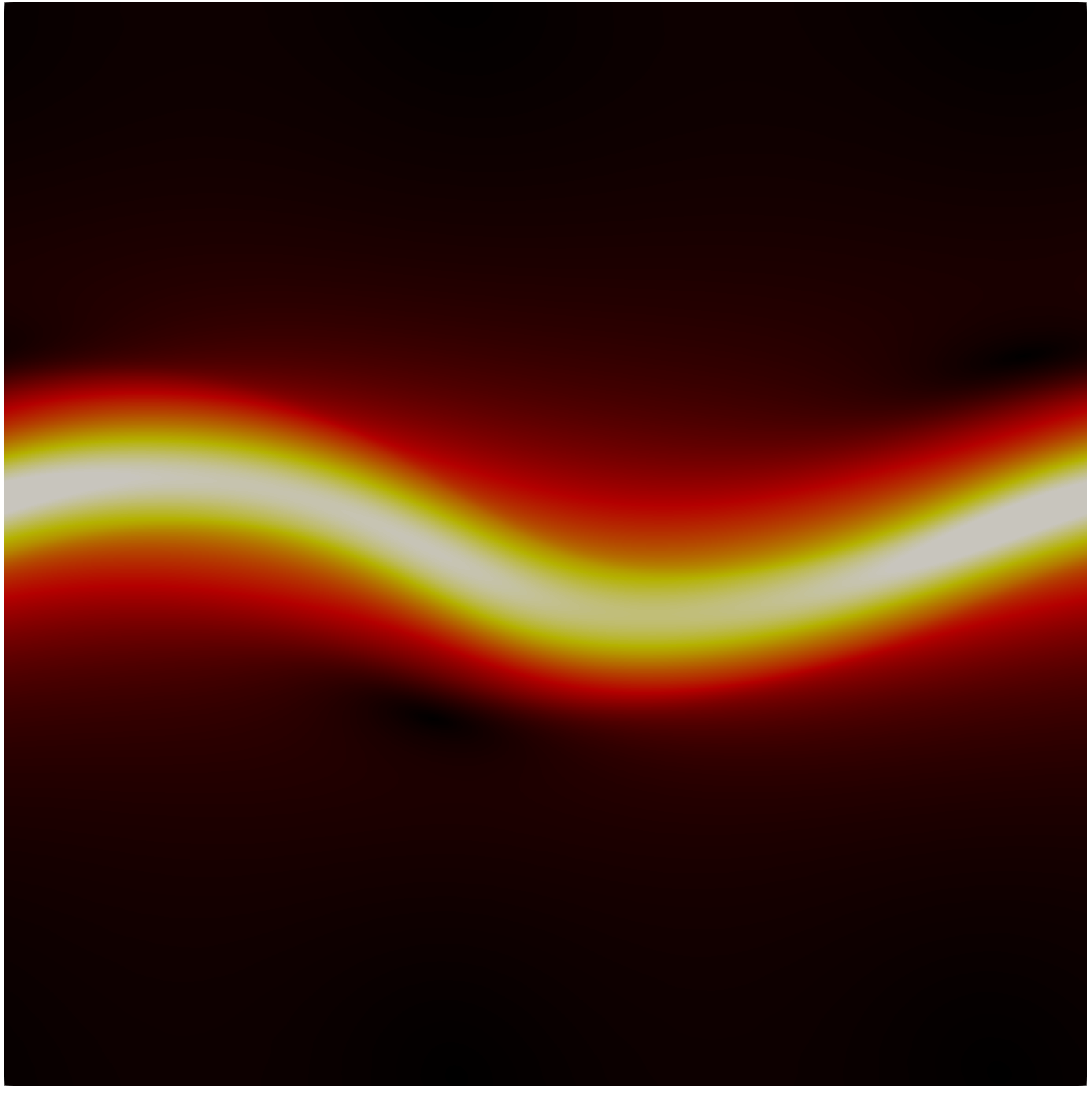}
		\caption{$t=24$}
		\label{fig:BJv24}
	\end{subfigure}
	\hfill
	\begin{subfigure}[b]{0.29\textwidth}
		\centering
		\includegraphics[width=\textwidth]{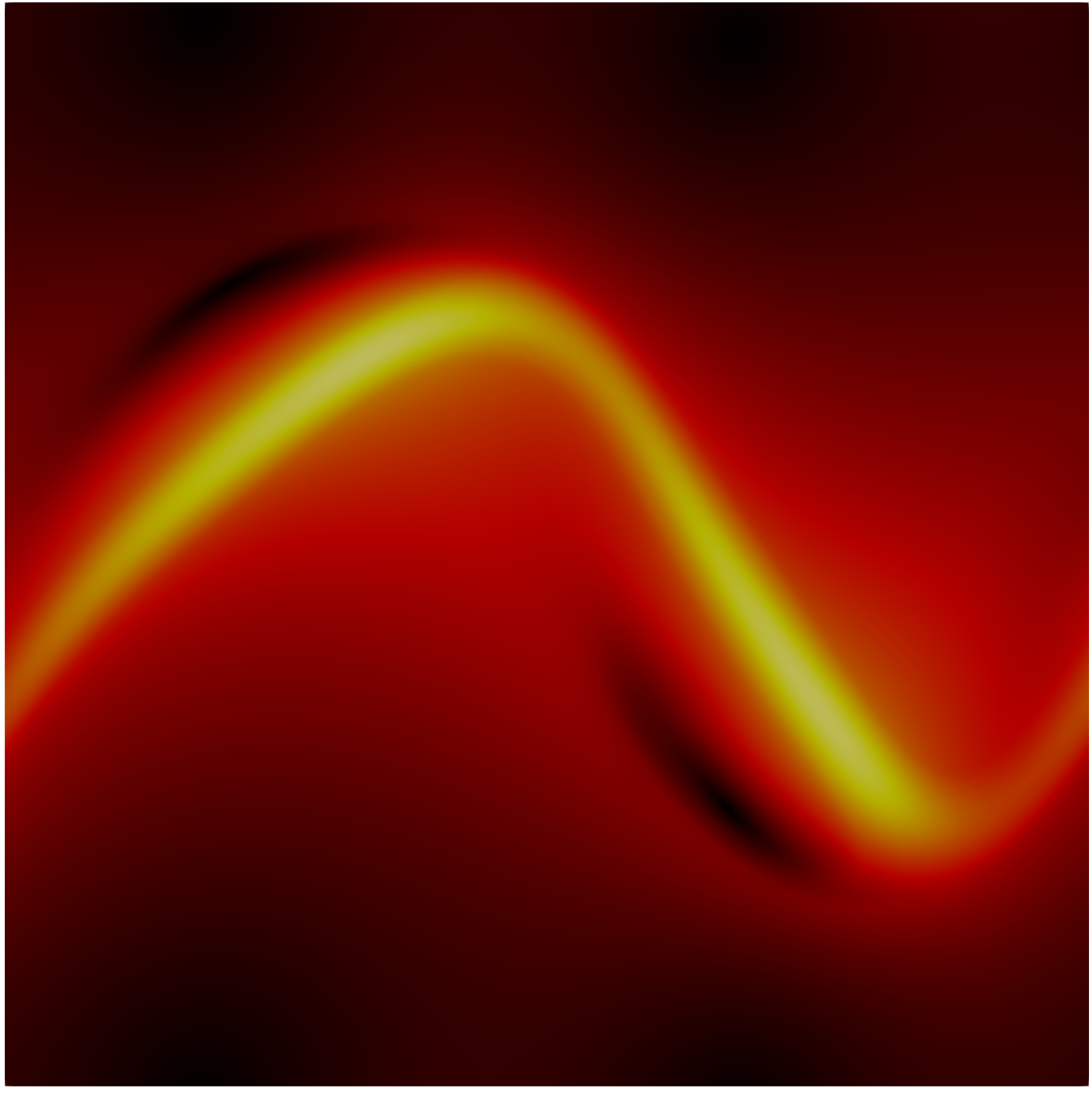}
		\caption{$t=36$}
		\label{fig:BJv36}
	\end{subfigure}
	\hfill
	\begin{subfigure}[b]{0.38\textwidth}
		\centering
		\includegraphics[width=\textwidth]{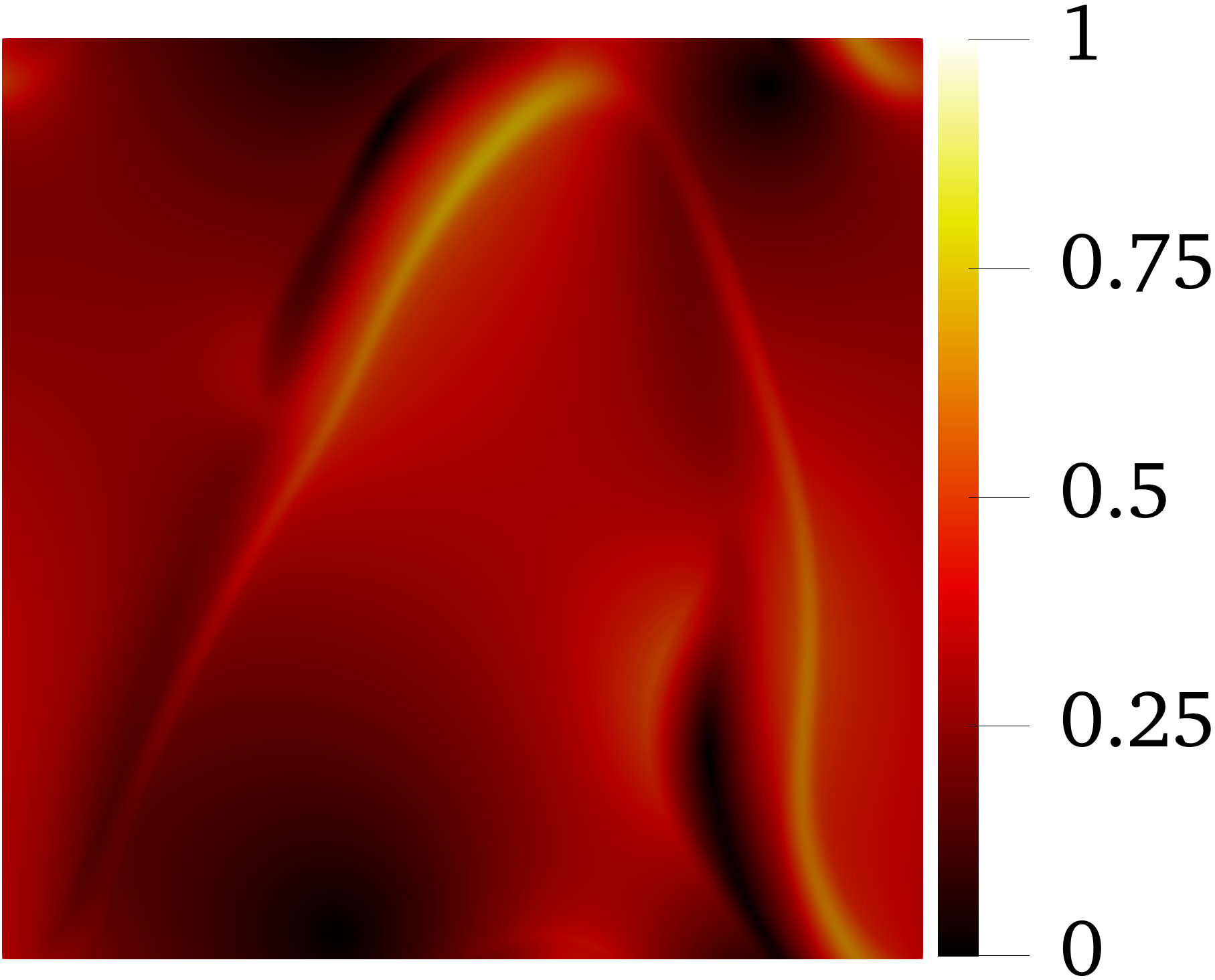}
		\caption{$t=48$}
		\label{fig:BJv48}
	\end{subfigure}
	\begin{subfigure}[b]{0.29\textwidth}
		\centering
		\includegraphics[width=\textwidth]{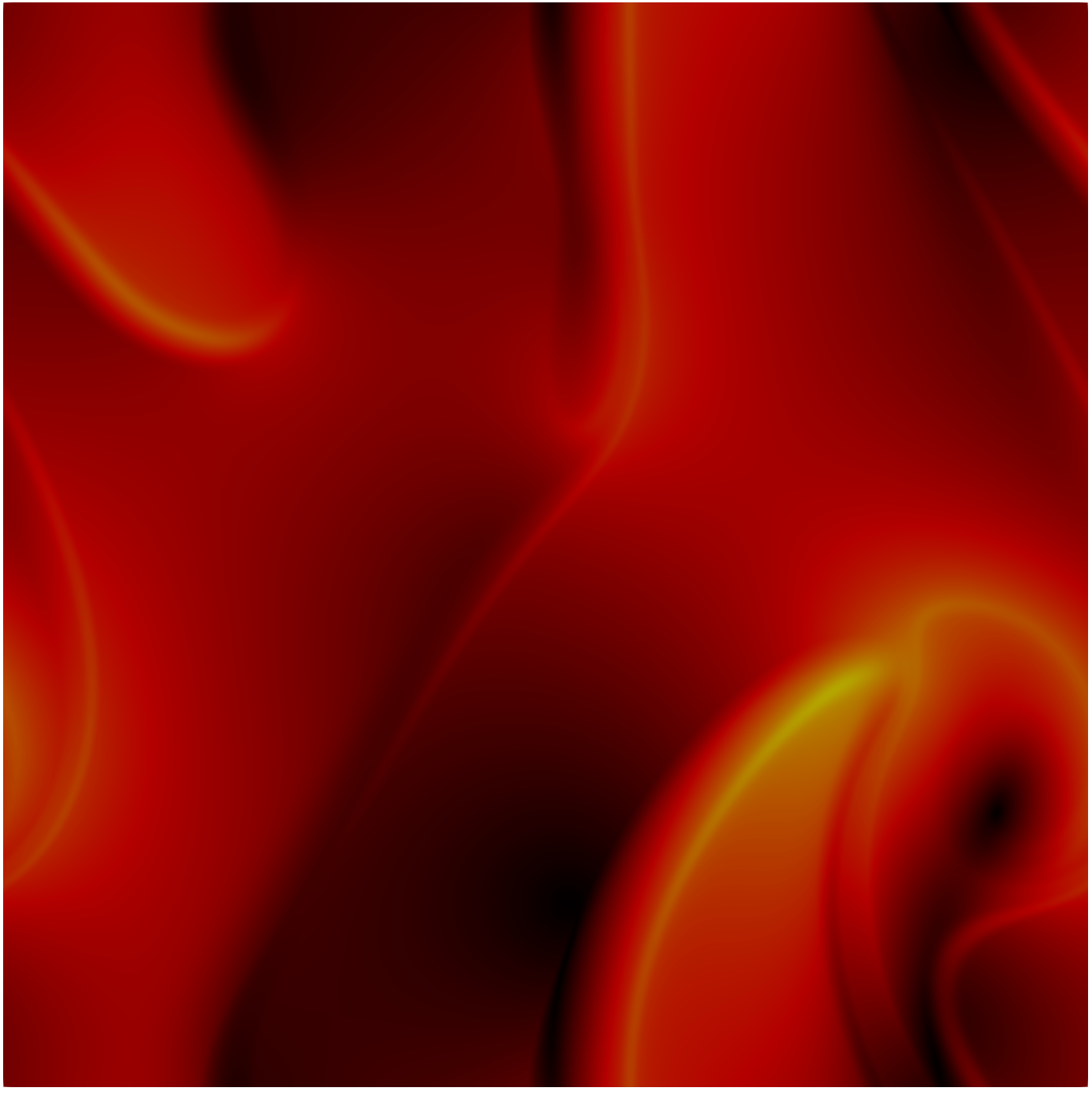}
		\caption{$t=60$}
		\label{fig:BJv60}
	\end{subfigure}
	\hfill
	\begin{subfigure}[b]{0.29\textwidth}
		\centering
		\includegraphics[width=\textwidth]{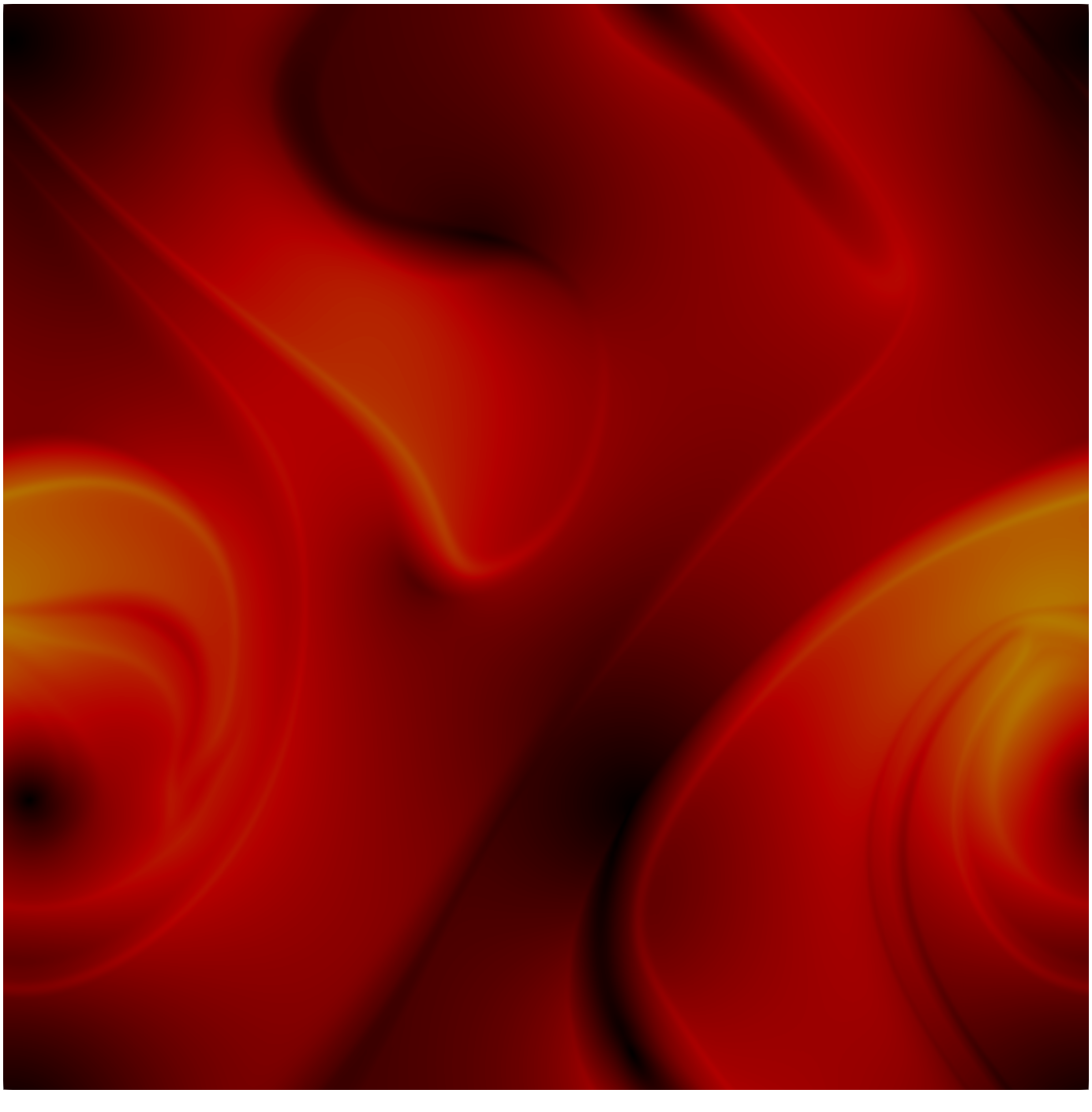}
		\caption{$t=72$}
		\label{fig:BJv72}
	\end{subfigure}
	\hfill
	\begin{subfigure}[b]{0.38\textwidth}
		\centering
		\includegraphics[width=\textwidth]{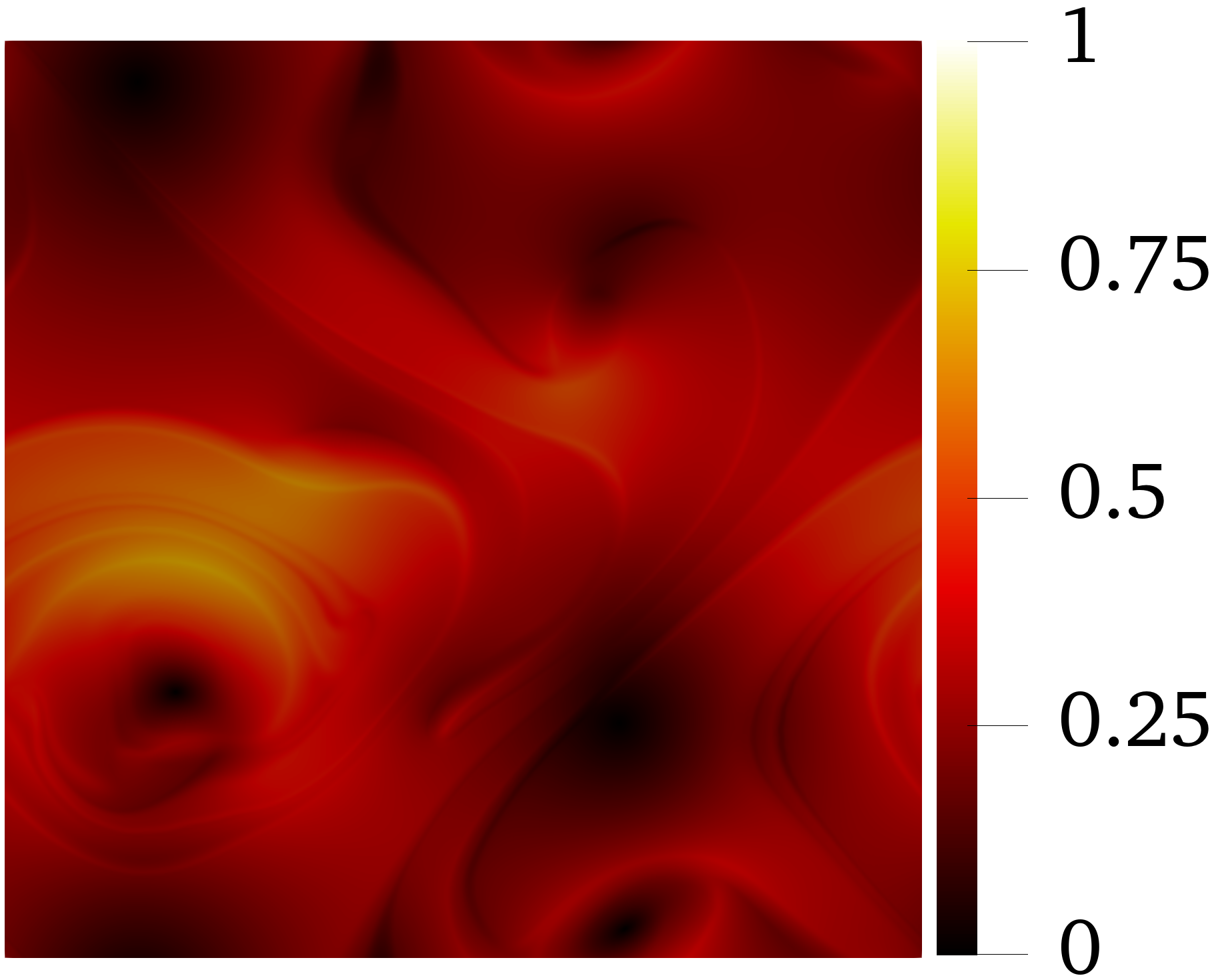}
		\caption{$t=84$}
		\label{fig:BJv84}
	\end{subfigure}
	\caption{Bickley jet -- Magnitude of the velocity field at different simulation times, computed with $p=4$ polynomials and a DIRK(3,3) temporal scheme.}
	\label{fig:BJmodel}
\end{figure}

A certain configuration of the Bickley jet case (see for instance \cite{Poulin2003}), using the shallow water equations, illustrates an example of the convection model.
The example models the temporal evolution of a jet flow in the square domain $\Omega = (-2\pi, 2\pi)^2$, subject to slight initial perturbations on the velocity field.
The problem is solved employing a Cartesian mesh of $128\times128$ quadrilaterals and periodic boundary conditions, given a non-dimensional gravity value of $g=10^4$.
The simulation employs $p=4$ polynomials and a DIRK(3,3) temporal integration scheme.
Some sketches of the velocity magnitude field for different simulation times are depicted in Figure~\ref{fig:BJmodel}. \texttt{Exasim} solves this case without need of any physical or artificial diffusion, showing a good resolution and propagation of the flow perturbations.

\subsection{Taylor-Green vortex}
Finally, the Taylor-Green vortex at $Re = 1600$ is considered to show the potential of \texttt{Exasim} in solving a 3D case featuring several millions of degrees of freedom. The case is solved using an implicit large-eddy simulation (ILES) approach, which relies on the inherent numerical dissipation of the DG discretization, contrary to introducing a subgrid scale (SGS) model to account for the small scales of the flow.
The example is solved in $\Omega = (0,2\pi)^3$ with periodic boundary conditions, employing a structured mesh of $64^3$ cubes and fourth-order polynomials.

\begin{figure}[H]
	\centering
	\begin{subfigure}[b]{0.48\textwidth}
		\centering
		\includegraphics[width=\textwidth]{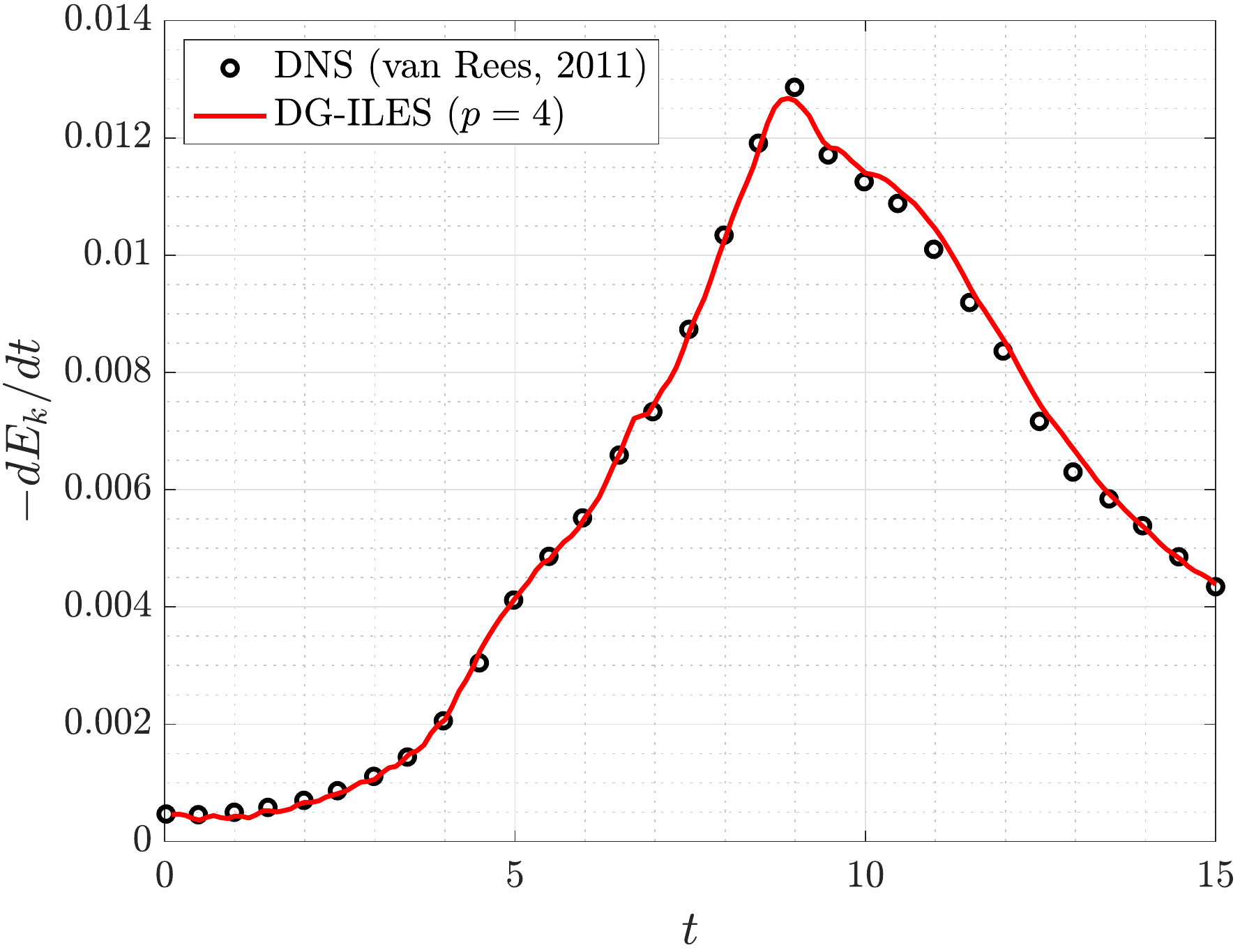}
		\caption{Kinetic energy rate}
		\label{fig:dEkdt}
		\vspace*{2pt}
	\end{subfigure}
	\hfill
	\begin{subfigure}[b]{0.48\textwidth}
		\centering
		\includegraphics[width=\textwidth]{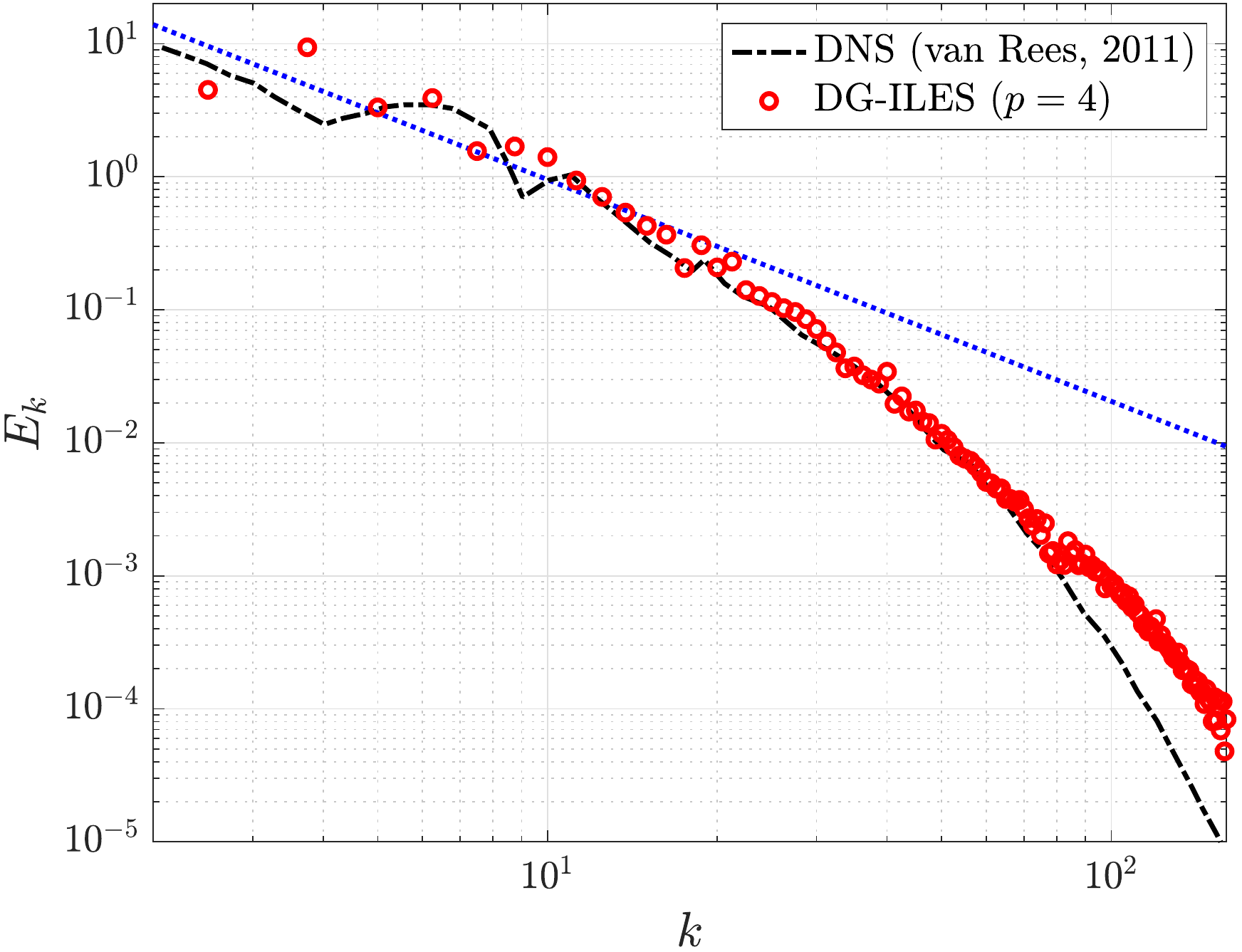}
		\caption{Kinetic energy spectrum at $t=9$.}
		\label{fig:Ek_k}
	\end{subfigure}
	\caption{Taylor-Green vortex -- Evolution of the kinetic energy rate with time (a) and kinetic energy spectrum at non-dimensional time $t=9$ (b). The dotted line indicates a slope of $-5/3$.}
	\label{fig:TGVEk}
\end{figure}

The kinetic energy dissipation is evaluated in Figure~\ref{fig:TGVEk} in comparison to a DNS reference solution~\cite{Rees2011}. On the one hand, Figure~\ref{fig:dEkdt} illustrates the evolution of the  kinetic energy rate, showing a strong agreement between the present ILES solution and the reference DNS solution.
On the other hand, Figure~\ref{fig:Ek_k} depicts the kinetic energy spectrum at the instant of maximum dissipation, $t=9$. 
The ILES solution establishes a close comparison with respect to the reference solution for wavenumbers $k<80$, whereas it overpredicts the energy in the highest wavenumbers. The wavespectrum also shows a first range of wavenumbers up to $k\simeq 20$ when it follows a power law with exponent close to the theoretical value of $-5/3$.

\section{Impact} \label{sc:impact}

\texttt{Exasim} is aimed at making DG methods accessible to users, offering at the same time a robust solver capable to handle large nonlinear systems of equations by means of innovative numerical algorithms.
To do so, a high-level interface allows users to specify the analytical expressions describing fluxes, source terms, boundary conditions and initial conditions of a general parametrized PDE system in simple Python, Julia or Matlab scripts.
This simple preprocessing step is then integrated within C++ and CUDA kernels with MPI-based parallelization, giving rise to a high-performance code capable to run on several machines, from laptops to supercomputers, with both CPU and GPU processors.

Furthermore, the software employs a set of GPU-accelerated numerical algorithms suited for this kind of architectures, allowing to take advantage of its computational power and to dramatically increase the scale and size of the numerical simulations \texttt{Exasim} is able to tackle.
In this manner, the software represents a key tool itself to face practical problems of interest in different physical applications whose computational demands are inaccessible by existing codes.

To this end, \texttt{Exasim} has been already used in different large-scale LES computations predicting transitional and turbulent flows in transonic or hypersonic applications \cite{Terrana2020,Nguyen2022}.
In addition, given the simplicity for formulating PDE models, \texttt{Exasim} has the potential to facilitate the introduction of modern high-order DG discretizations in physical systems involving complex descriptions and coupled phenomena.

Finally, the modular structure of \texttt{Exasim} permits to exploit the full capabilities of external libraries and packages.
For instance, this is the case of Enzyme, which has been integrated within the software to perform forward mode automatic differentiation in the matrix-vector products associated to the GMRES iterations.

\section{Conclusions} \label{sc:conclusions}

This paper presents \texttt{Exasim}, an open-source software for generating discontinuous Galerkin codes for the numerical solution of PDEs.
The code features a high-level user interface in Julia, Python and Matlab for preprocessing and code generation and combines it with C++ and CUDA/MPI kernels, producing a high-performance code able to run on CPU and GPU platforms.
The code exploits a matrix-free solution method that provides full GPU functionality and excellent scalability on this kind of architectures. The software has been validated on many different applications and examples, and it constitutes both a basic learning framework and an innovative numerical tool for advanced computational research used to improve our understanding of complex flow physics.

\section{Conflict of Interest}
The authors declare that there are no known conflicts of interest associated with this publication and that there has been no significant financial support for this work that could have influenced its outcome.

\section*{Acknowledgements}
The authors would like to thank Sebastien Terrana his contribution in the development of the software, as well as William Moses for his assistance in setting up Enzyme within \texttt{Exasim}.
We  gratefully acknowledge the United States  Department of Energy (under contract DE-NA0003965)  and the National Science Foundation (under grant number NSF-PHY-2028125) for supporting this work. Finally, the authors thank the Oak Ridge Leadership Computing Facility for providing access to their GPU clusters.



\bibliographystyle{elsarticle-num} 
\bibliography{library}

\begin{thebibliography}{10}
\expandafter\ifx\csname url\endcsname\relax
  \def\url#1{\texttt{#1}}\fi
\expandafter\ifx\csname urlprefix\endcsname\relax\def\urlprefix{URL }\fi
\expandafter\ifx\csname href\endcsname\relax
  \def\href#1#2{#2} \def\path#1{#1}\fi

\bibitem{Kroll2015}
N.~Kroll, C.~Hirsch, F.~Bassi, C.~Johnston, K.~Hillewaert (Eds.), {IDIHOM}:
  Industrialization of High-Order Methods - A Top-Down Approach, Springer
  International Publishing, 2015.

\bibitem{Kroll2009}
N.~Kroll, ADIGMA: A European Project on the Development of Adaptive Higher
  Order Variational Methods for Aerospace Applications, Aerospace Sciences
  Meetings, American Institute of Aeronautics and Astronautics, 2009.

\bibitem{Wang2013}
Z.~J. Wang, K.~Fidkowski, R.~Abgrall, F.~Bassi, D.~Caraeni, A.~Cary,
  H.~Deconinck, R.~Hartmann, K.~Hillewaert, H.~T. Huynh, N.~Kroll, G.~May,
  P.-O. Persson, B.~van Leer, M.~Visbal, High-order {CFD} methods: current
  status and perspective, Int. J. Numer. Methods Fluids 72~(8) (2013) 811--845.

\bibitem{Slotnick2014}
J.~Slotnick, A.~Khodadoust, J.~Alonso, D.~L. Darmofal, W.~Gropp, E.~Lurie,
  D.~J. Mavriplis, {CFD vision 2030 study : A path to revolutionary
  computational aerosciences}, Tech. Rep. NASA-CR-2014-218178, NASA (2014).

\bibitem{Ekaterinaris2005}
J.~A. Ekaterinaris, High-order accurate, low numerical diffusion methods for
  aerodynamics, Prog. Aerosp. Sci. 41~(3-4) (2005) 192--300.

\bibitem{Cockburn-CKS:2000}
B.~Cockburn, G.~E. Karniadakis, C.-W. Shu, The development of discontinuous
  {G}alerkin methods, in: Discontinuous {G}alerkin {M}ethods, Springer-Verlag
  Berlin Heidelberg, Berlin, Germany, 2000, pp. 3--50.

\bibitem{Hesthaven2010}
T.~W. Jan S.~Hesthaven, Nodal Discontinuous Galerkin Methods, Springer New
  York, 2010.

\bibitem{Cockburn-CLS:1989}
B.~Cockburn, S.-Y. Lin, C.-W. Shu, {TVB} {R}unge-{K}utta local projection
  discontinuous {G}alerkin finite element method for conservation laws {III}:
  one-dimensional systems, J. Comput. Phys. 84~(1) (1989) 90--113.

\bibitem{Cockburn-CS:1998}
B.~Cockburn, C.-W. Shu, The {R}unge{\textendash}{K}utta {D}iscontinuous
  {G}alerkin {M}ethod for {C}onservation {L}aws {V}, J. Comput. Phys. 141~(2)
  (1998) 199--224.

\bibitem{Bassi-BR:97}
F.~Bassi, S.~Rebay, High-order accurate discontinuous finite element solution
  of the 2{D} {E}uler equations, J. Comput. Phys. 138~(2) (1997) 251--285.

\bibitem{Bassi-BR:1997b}
F.~Bassi, S.~Rebay, A high-order accurate discontinuous finite element method
  for the numerical solution of the compressible {N}avier-{S}tokes equations,
  J. Comput. Phys. 131~(2) (1997) 267--279.

\bibitem{Bassi-BR:2002}
F.~Bassi, S.~Rebay, Numerical evaluation of two discontinuous {G}alerkin
  methods for the compressible {N}avier-{S}tokes equations, Int. J. Numer.
  Methods Fluids 40~(1-2) (2002) 197--207.

\bibitem{HartmannHouston:2003}
R.~Hartmann, P.~Houston, Adaptive discontinuous {G}alerkin finite element
  methods for nonlinear hyperbolic conservation laws, {SIAM} Journal on
  Scientific Computing 24~(3) (2003) 979--1004.

\bibitem{Balan-BWM:2015}
A.~Balan, M.~Woopen, G.~May, Hp-adaptivity on anisotropic meshes for hybridized
  discontinuous {G}alerkin scheme, AIAA paper (jun 2015).

\bibitem{Giorgiani-GFH:2014}
G.~Giorgiani, S.~Fern{\'{a}}ndez-M{\'{e}}ndez, A.~Huerta, Hybridizable
  discontinuous {G}alerkin with degree adaptivity for the incompressible
  {N}avier-{S}tokes equations, Comput. Fluids 98 (2014) 196--208.

\bibitem{Cangiani2017}
A.~Cangiani, Z.~Dong, E.~H. Georgoulis, P.~Houston, $hp$-{V}ersion
  Discontinuous {G}alerkin Methods on Polygonal and Polyhedral Meshes,
  Springer-Verlag, 2017.

\bibitem{Roca-RNP:2013}
X.~Roca, C.~Nguyen, J.~Peraire, Scalable parallelization of the hybridized
  discontinuous {G}alerkin method for compressible flow, AIAA Paper (jun 2013).

\bibitem{Cockburn-CG:2004}
B.~Cockburn, J.~Gopalakrishnan, A characterization of hybridized mixed methods
  for second order elliptic problems, SIAM Journal on Numerical Analysis 42~(1)
  (2004) 283--301.

\bibitem{Cockburn2009}
B.~Cockburn, J.~Guzm{\'{a}}n, S.-C. Soon, H.~K. Stolarski, {An Analysis of the
  Embedded Discontinuous Galerkin Method for Second-Order Elliptic Problems},
  SIAM Journal on Numerical Analysis 47~(4) (2009) 2686--2707.

\bibitem{Nguyen2015c}
N.~C. Nguyen, J.~Peraire, B.~Cockburn, {A class of embedded discontinuous
  Galerkin methods for computational fluid dynamics}, Journal of Computational
  Physics 302 (2015) 674--692.

\bibitem{CockburnShu1998}
B.~Cockburn, C.-W. Shu, {The local discontinuous Galerkin method for
  time-dependent convection-diffusion systems}, SIAM Journal on Numerical
  Analysis 35~(6) (1998) 2440--2463.

\bibitem{Cockburn-CNP:2010}
B.~Cockburn, N.~C. Nguyen, J.~Peraire, A comparison of {HDG} methods for
  {S}tokes flow, J. Sci. Comput. 45~(1-3) (2010) 215--237.

\bibitem{PerairePersson08}
J.~Peraire, P.-O. Persson, {The compact discontinuous Galerkin method for
  elliptic problems}, SIAM Journal on Scientific Computing 30~(4) (2008)
  1806--1824.

\bibitem{VilaPerez2020}
J.~Vila-P{\'{e}}rez, M.~Giacomini, R.~Sevilla, A.~Huerta, {Hybridisable
  Discontinuous Galerkin Formulation of Compressible Flows}, Archives of
  Computational Methods in Engineering 28~(2) (2020) 753--784.

\bibitem{RS-SH:16}
R.~Sevilla, A.~Huerta, Tutorial on {H}ybridizable {D}iscontinuous {G}alerkin
  ({HDG}) for second-order elliptic problems, in: J.~Schr{\"o}der, P.~Wriggers
  (Eds.), Advanced Finite Element Technologies, Vol. 566 of CISM International
  Centre for Mechanical Sciences, Springer International Publishing, 2016, pp.
  105--129.

\bibitem{Ciuca2020}
C.~Ciuc{\u{a}}, P.~Fernandez, A.~Christophe, N.~Nguyen, J.~Peraire, Implicit
  hybridized discontinuous {G}alerkin methods for compressible
  magnetohydrodynamics, Journal of Computational Physics: X 5 (2020) 100042.

\bibitem{Fernandez2018a}
P.~Fernandez, A.~Christophe, S.~Terrana, N.~C. Nguyen, J.~Peraire, {Hybridized
  Discontinuous Galerkin Methods for Wave Propagation}, Journal of Scientific
  Computing 77~(3) (2018) 1566--1604.

\bibitem{Fernandez2017a}
P.~Fernandez, N.~C. Nguyen, J.~Peraire, {The hybridized discontinuous Galerkin
  method for implicit large-eddy simulation of transitional turbulent flows},
  Journal of Computational Physics 336 (2017) 308--329.

\bibitem{Nguyen2012}
N.~C. Nguyen, J.~Peraire, {Hybridizable discontinuous Galerkin methods for
  partial differential equations in continuum mechanics}, Journal of
  Computational Physics 231~(18) (2012) 5955--5988.

\bibitem{Nguyen-NPC:20011acoustics}
N.~Nguyen, J.~Peraire, B.~Cockburn, High-order implicit hybridizable
  discontinuous galerkin methods for acoustics and elastodynamics, Journal of
  Computational Physics 230~(10) (2011) 3695--3718.

\bibitem{Giacomini-GKSH:2018}
M.~Giacomini, A.~Karkoulias, R.~Sevilla, A.~Huerta, A superconvergent {HDG}
  method for {S}tokes flow with strongly enforced symmetry of the stress
  tensor, J. Sci. Comput. 77~(3) (2018) 1679--1702.

\bibitem{Sevilla-SGKH:2018}
R.~Sevilla, M.~Giacomini, A.~Karkoulias, A.~Huerta, A superconvergent
  hybridisable discontinuous {G}alerkin method for linear elasticity, Int. J.
  Numer. Methods Eng. 116~(2) (2018) 91--116.

\bibitem{VidalCodina2018}
F.~Vidal-Codina, N.~Nguyen, S.-H. Oh, J.~Peraire, A hybridizable discontinuous
  galerkin method for computing nonlocal electromagnetic effects in
  three-dimensional metallic nanostructures, Journal of Computational Physics
  355 (2018) 548--565.

\bibitem{VidalCodina2021}
F.~Vidal-Codina, N.-C. Nguyen, C.~Cirac{\`{\i}}, S.-H. Oh, J.~Peraire, A nested
  hybridizable discontinuous galerkin method for computing second-harmonic
  generation in three-dimensional metallic nanostructures, Journal of
  Computational Physics 429 (2021) 110000.

\bibitem{Huerta-HARP:2013}
A.~Huerta, A.~Angeloski, X.~Roca, J.~Peraire, Efficiency of high-order elements
  for continuous and discontinuous {G}alerkin methods, Int. J. Numer. Methods
  Eng. 96~(9) (2013) 529--560.

\bibitem{May-WBMS-14}
M.~Woopen, A.~Balan, G.~May, J.~Sch\"{u}tz, A comparison of hybridized and
  standard {DG} methods for target-based hp-adaptive simulation of compressible
  flow, Comput. Fluids 98 (2014) 3 -- 16.

\bibitem{Kronbichler2019}
M.~Kronbichler, K.~Kormann, W.~A. Wall, Fast matrix-free evaluation of
  hybridizable discontinuous {G}alerkin operators, in: Lecture Notes in
  Computational Science and Engineering, Springer International Publishing,
  2019, pp. 581--589.

\bibitem{Terrana2020}
S.~Terrana, N.~C. Nguyen, J.~Peraire, {GPU-accelerated Large Eddy Simulation of
  Hypersonic Flows}, in: AIAA Scitech 2020 Forum, 2020, pp. AIAA--2020--1062.

\bibitem{CockburnKanschatSchoetzauNS05}
B.~Cockburn, G.~Kanschat, D.~Sch{\"{o}}tzau, {A locally conservative
  {\{}LDG{\}} method for the incompressible {\{}N{\}}avier-{\{}S{\}}tokes
  equations}, Math. Comp. 74 (2005) 1067--1095.

\bibitem{Nguyen2022}
N.~C. Nguyen, S.~Terrana, J.~Peraire, Large-eddy simulation of transonic buffet
  using matrix-free discontinuous galerkin method, {AIAA} Journal (2022) 1--18.

\bibitem{ExaDG2020}
D.~Arndt, N.~Fehn, G.~Kanschat, K.~Kormann, M.~Kronbichler, P.~Munch, W.~A.
  Wall, J.~Witte, {ExaDG}: High-order discontinuous {G}alerkin for the
  exa-scale, in: H.-J. Bungartz, S.~Reiz, B.~Uekermann, P.~Neumann, W.~E. Nagel
  (Eds.), Software for Exascale Computing - SPPEXA 2016-2019, Springer
  International Publishing, Cham, 2020, pp. 189--224.

\bibitem{Arndt2021}
D.~Arndt, W.~Bangerth, D.~Davydov, T.~Heister, L.~Heltai, M.~Kronbichler,
  M.~Maier, J.-P. Pelteret, B.~Turcksin, D.~Wells, The deal.{II} finite element
  library: Design, features, and insights, Computers {\&} Mathematics with
  Applications 81 (2021) 407--422.

\bibitem{Anderson2021}
R.~Anderson, J.~Andrej, A.~Barker, J.~Bramwell, J.-S. Camier, J.~Cerveny,
  V.~Dobrev, Y.~Dudouit, A.~Fisher, T.~Kolev, W.~Pazner, M.~Stowell, V.~Tomov,
  I.~Akkerman, J.~Dahm, D.~Medina, S.~Zampini, {MFEM}: A modular finite element
  methods library, Computers {\&} Mathematics with Applications 81 (2021)
  42--74.

\bibitem{flexi_general}
F.~Hindenlang, G.~J. Gassner, C.~Altmann, A.~Beck, M.~Staudenmaier, C.-D. Munz,
  Explicit discontinuous galerkin methods for unsteady problems, Computers \&
  Fluids 61 (2012) 86--93.

\bibitem{Cantwell2015}
C.~Cantwell, D.~Moxey, A.~Comerford, A.~Bolis, G.~Rocco, G.~Mengaldo, D.~D.
  Grazia, S.~Yakovlev, J.-E. Lombard, D.~Ekelschot, B.~Jordi, H.~Xu,
  Y.~Mohamied, C.~Eskilsson, B.~Nelson, P.~Vos, C.~Biotto, R.~Kirby,
  S.~Sherwin, Nektar++: An open-source spectral/hp element framework, Computer
  Physics Communications 192 (2015) 205--219.

\bibitem{Ching2022}
E.~J. Ching, B.~Bornhoft, A.~Lasemi, M.~Ihme, Quail: A lightweight open-source
  discontinuous galerkin code in python for teaching and prototyping,
  {SoftwareX} 17 (2022) 100982.

\bibitem{Giacomini2020}
M.~Giacomini, R.~Sevilla, A.~Huerta, {{HDGlab}: An Open-Source Implementation
  of the Hybridisable Discontinuous Galerkin Method in {MATLAB}}, Archives of
  Computational Methods in Engineering 28~(3) (2020) 1941--1986.

\bibitem{Dedner2010}
A.~Dedner, R.~Klöfkorn, M.~Nolte, M.~Ohlberger, A generic interface for
  parallel and adaptive discretization schemes: abstraction principles and the
  dune-fem module, Computing 90~(3-4) (2010) 165--196.

\bibitem{NGSolve}
J.~Sch\"{o}berl, C++11 implementation of finite elements in ngsolve, Tech.
  rep., Institute for Analysis and Scientific Computing, Vienna University of
  Technology - TU Wien (2014).

\bibitem{Prudhomme2021}
C.~Prud'homme, V.~Chabannes, {StephaneVeys}, A.~Ancel, T.~Metivet, R.~Hild,
  {Jbwahl}, C.~Daversin-Catty, G.~Dollé, {Tarabay}, {Lsala}, {LANTZT},
  {Doyeux}, {Trophime}, {Abdoulaye SAMAKE}, B.~Vanthong, M.~ISMAIL, V.~Huber,
  {Kyoshe Winstone}, {Schenone}, T.~Saigre, {, Philippe}, D.~Prada, {Lberti},
  {Prj-}, D.~Barbier, {Clayrc}, {, Yacine}, J.~Veysset, feelpp/feelpp: Feel++
  v109 (2021).

\bibitem{Reuter2020}
B.~Reuter, {FESTUNG}: A {MATLAB} / {GNU} octave toolbox for the discontinuous
  galerkin method. part {IV}: Generic problem framework and model-coupling
  interface, Communications in Computational Physics 28~(2) (2020) 827--876.

\bibitem{Kloeckner2012}
A.~Klöckner, T.~Warburton, J.~S. Hesthaven, Solving wave equations on
  unstructured geometries, in: {GPU} Computing Gems Jade Edition, Elsevier,
  2012, pp. 225--242.

\bibitem{Witherden2014}
F.~Witherden, A.~Farrington, P.~Vincent, {PyFR}: An open source framework for
  solving advection{\textendash}diffusion type problems on streaming
  architectures using the flux reconstruction approach, Computer Physics
  Communications 185~(11) (2014) 3028--3040.

\bibitem{Alexander77}
R.~Alexander, {Diagonally implicit Runge-Kutta methods for stiff ODEs}, SIAM J.
  Numer. Anal. 14 (1977) 1006--1021.

\bibitem{Geuzaine2009}
C.~Geuzaine, J.-F. Remacle, Gmsh: A 3-d finite element mesh generator with
  built-in pre- and post-processing facilities, International Journal for
  Numerical Methods in Engineering 79~(11) (2009) 1309--1331.

\bibitem{CUBIT}
M.~Skroch, S.~J. Owen, M.~L. Staten, R.~W. Quadros, B.~Hanks, B.~Clark,
  T.~Hensley, C.~Ernst, R.~Morris, C.~McBride, C.~Stimpson, J.~Perry,
  M.~Richardson, K.~Merkley, {CUBIT: Geometry and Mesh Generation Toolkit.
  16.02 User Documentation}, Tech. rep., Sandia National Laboratories (2022).

\bibitem{CGAL}
{The CGAL Project}, {CGAL} User and Reference Manual, {5.4} Edition, {CGAL
  Editorial Board}, 2022.

\bibitem{Persson2004}
P.-O. Persson, G.~Strang, A simple mesh generator in {MATLAB}, {SIAM} Review
  46~(2) (2004) 329--345.

\bibitem{Si2015}
H.~Si, {{TetGen}, a Delaunay-Based Quality Tetrahedral Mesh Generator}, {ACM}
  Transactions on Mathematical Software 41~(2) (2015) 1--36.

\bibitem{Dapogny2014}
C.~Dapogny, C.~Dobrzynski, P.~Frey, Three-dimensional adaptive domain
  remeshing, implicit domain meshing, and applications to free and moving
  boundary problems, Journal of Computational Physics 262 (2014) 358--378.

\bibitem{Cignoni2008}
P.~Cignoni, M.~Callieri, M.~Corsini, M.~Dellepiane, F.~Ganovelli, G.~Ranzuglia,
  Meshlab: an open-source mesh processing tool (2008).

\bibitem{Nicolas2013}
G.~Nicolas, T.~Fouquet, Adaptive mesh refinement for conformal
  hexahedralmeshes, Finite Elements in Analysis and Design 67 (2013) 1--12.

\bibitem{Persson2009a}
P.-O. Persson, J.~Peraire, {Curved mesh generation and mesh refinement using
  Lagrangian solid mechanics}, in: 47th AIAA Aerospace Sciences Meeting, no.
  949, 2009.

\bibitem{Gargallo-Peiro2015}
A.~Gargallo-Peir{\'{o}}, X.~Roca, J.~Peraire, J.~Sarrate, {Optimization of a
  regularized distortion measure to generate curved high-order unstructured
  tetrahedral meshes}, International Journal for Numerical Methods in
  Engineering 103~(5) (2015) 342--363.

\bibitem{Moses2020}
W.~Moses, V.~Churavy, Instead of rewriting foreign code for machine learning,
  automatically synthesize fast gradients, in: H.~Larochelle, M.~Ranzato,
  R.~Hadsell, M.~F. Balcan, H.~Lin (Eds.), Advances in Neural Information
  Processing Systems, Vol.~33, Curran Associates, Inc., 2020, pp. 12472--12485.

\bibitem{Moses2021}
W.~S. Moses, V.~Churavy, L.~Paehler, J.~Hückelheim, S.~H.~K. Narayanan,
  M.~Schanen, J.~Doerfert, Reverse-mode automatic differentiation and
  optimization of {GPU} kernels via {E}nzyme, in: Proceedings of the
  International Conference for High Performance Computing, Networking, Storage
  and Analysis, {ACM}, 2021.

\bibitem{Hader2019}
C.~Hader, H.~F. Fasel, {Direct numerical simulations of hypersonic
  boundary-layer transition for a flared cone: Fundamental breakdown}, Journal
  of Fluid Mechanics 869 (2019) 341--384.

\bibitem{Chynoweth2019}
B.~C. Chynoweth, S.~P. Schneider, C.~Hader, H.~Fasel, A.~Batista, J.~Kuehl,
  T.~J. Juliano, B.~M. Wheaton, {History and progress of boundary-layer
  transition on a Mach-6 flared cone}, Journal of Spacecraft and Rockets 56~(2)
  (2019) 333--346.

\bibitem{Baumeister1994}
K.~J. Baumeister, K.~L. Kreider, Scattering cross section of sound waves by the
  modal element method, in: Winter Annual Meeting by the American Society of
  Mechanical Engineers, 1994.

\bibitem{Poulin2003}
F.~J. Poulin, G.~R. Flierl, The nonlinear evolution of barotropically unstable
  jets, Journal of Physical Oceanography 33~(10) (2003) 2173--2192.

\bibitem{Rees2011}
W.~M. van Rees, A.~Leonard, D.~Pullin, P.~Koumoutsakos, A comparison of vortex
  and pseudo-spectral methods for the simulation of periodic vortical flows at
  high {R}eynolds numbers, Journal of Computational Physics 230~(8) (2011)
  2794--2805.

\end{thebibliography}

\end{document}